\documentstyle[12pt,aasms4]{article}
\def\et{{\it et al.}}
\def\cbr{\mbox{\scriptsize CMBR}}
\def\gap{\stackrel{>}{_\sim}}
\def\lap{\stackrel{<}{_\sim}}
\def\tcbr{$T_{\cbr}$}

\begin{document}
\title{An Absolute Measurement of the Cosmic Microwave 
Background Radiation Temperature at 10.7~GHz}
\author{S. T. Staggs,
	N. C. Jarosik,\altaffilmark{1} S. S. Meyer, 
	D. T. Wilkinson\altaffilmark{1}  }

\affil{Enrico Fermi Institute\\
University of Chicago, Chicago, IL 60637.}

\altaffiltext{1}{Joseph Henry Laboratories and Physics
Department, Princeton University, Princeton, NJ 08544}

\begin{abstract}

A balloon-borne experiment 
has measured the absolute temperature of 
the cosmic microwave background radiation (CMBR)
at 10.7~GHz to be 
\tcbr$=2.730\pm 0.014$~K.  The
error is the quadratic sum of several  systematic errors, with
statistical error of less than 0.1~mK. The instrument
comprises a
cooled corrugated horn antenna coupled to a total-power radiometer.
A cryogenic mechanical waveguide switch alternately connects the
radiometer to the horn and to an internal reference load. The
small measured temperature difference ($\lap 20$~mK) between
the sky signal and the reference load in conjunction with the use
of a cold front end keeps systematic instrumental corrections
small. Atmospheric and window emission are minimized by flying
the instrument at 24~km altitude.  A large outer ground screen and
smaller inner screen shield the instrument from stray radiation 
from the ground and the balloon.  In-flight tests 
constrain the magnitude of ground radiation
contamination, and low level interference is monitored through
observations in several narrow  frequency bands.

\end{abstract}

\keywords{cosmology:cosmic background radiation --- 
  cosmology: observations} 

\section{INTRODUCTION} 
\nopagebreak 

The residual radiation from the early universe retains a nearly 
blackbody spectrum today, allowing for its  interpretation as
a primeval remnant.   While the present spatial anisotropy of the
cosmic microwave background radiation (CMBR) maps out the
positions of small inhomogeneities in the early universe at the
epoch of decoupling, spectral distortions in the CMBR 
carry an imprint of the energetic history of the early 
universe. Processes releasing energy at redshift $z<10^{7}$ (see
\cite{Wri94} for examples) can cause
distortions in the CMBR spectrum, either by directly altering the
photon occupation number or by heating the matter  which may then
couple to the radiation.  Several frequency-dependent processes,
including nonradiative and radiative Compton
scattering and free-free emission, slowly relax the radiation
back toward a  blackbody spectrum; thus the frequency signature
of any distortion  provides clues to both the nature and epoch of
its origin. Smoot \& Scott~1996 give a review of distortions and
recent experimental results.  
The parameter $Y_{ff}$ describes distortions at long
wavelengths arising from free-free emission during an epoch of
re-ionization, while $y$ parameterizes Compton distortions (most
noticeable at short wavelengths) which have not yet undergone
much relaxation.  The chemical potential $\mu$ parameterizes
distortions in which the CMBR spectrum has relaxed to kinetic
equilibrium, but lacks the number of photons needed for a
blackbody spectrum.

The FIRAS instrument on the COBE satellite measured the CMBR
spectrum between 60~GHz and 630~GHz very precisely
(\cite{Fix96}), determining~\tcbr\ to be $2.728\pm .004$~K
(95\% CL) and constraining $y<1.5 \times 10^{-5}$ and 
$\mu<9 \times 10^{-5}$.     
At longer wavelengths, existing measurements are
substantially less accurate. All extant data taken together 
constrain $Y_{ff}< 1.9 \times 10^{-5}$ (\cite{Smo96}).  
Most progress on measurements
of or limits on $\mu$ and $Y_{ff}$ will now come from improved
longer wavelength data. This paper presents a recent accurate
measurement of~\tcbr\ at 10.7~GHz.

\section{EXPERIMENT DESIGN}
\nopagebreak

The balloon-borne apparatus (Figure 1)  features a
cold ($\approx 5$~K) corrugated horn  antenna with a  $15^\circ$
FWHM beam coupled through a cold waveguide switch to a total
power radiometer. The switch alternately connects the radiometer
to the antenna signal and to an internal temperature-regulated
reference load, consisting of iron-loaded epoxy absorber
thermally sunk inside a 2.8~kg copper waveguide block.  The
load's measured reflection is $< -30$~dB.  The
switch period is 85  seconds.  The horn antenna views the sky
through a thin ($0.001''$)  polypropylene window which separates
the dewar vacuum from the ambient pressure at float ($\approx
20$~torr).\footnote{The  window is too thin to withstand
atmospheric pressure; a metal plate seals off the dewar vacuum on
the ground and is pulled out of the way in flight.}  The input of
the radiometer incorporates a cryogenic circulator and a second
temperature-regulated load to mitigate errors associated with
small differences in the reflection coefficients of the horn and
reference load.  The temperatures of the two loads are measured
with calibrated germanium resistance thermometers (Lake Shore
model GR-200A-2500) and read out with precision 
AC bridges. During the flight these temperatures were stable to better than
$0.1$~mK over time scales of several minutes.  Additional 
temperature regulators maintained the switch 
and the circulator body within  $\approx 100$~mK of the sky signal
during flight.  

The signal is amplified in a cryogenic 9--12 GHz HEMT amplifier
supplied by the NRAO.
Subsequently, the signal bandpass is limited to 10--11~GHz in
a  fixed room-temperature filter. Two fixed filters
and  two tunable filters divide the signal into four frequency
bands.  One  linear polarization is detected. During the flight,
one of the tunable filters (with 50~MHz bandwidth) was
continuously stepped through the 1 GHz bandpass to monitor
possible radiofrequency interference; none was seen.   
The system noise temperature was $12~$K.
This paper reports the
results from the two fixed filters, with bandpasses $10.625\pm
.025$~GHz (Channel 2) and $10.69\pm.01$~GHz (Channel 3).

The gondola hangs $\approx 1100'$ below the balloon, which has a
diameter of $\approx 175'$ at float.  The gondola rotated about
once a minute for much of the flight, so that the instrument
observed a broad swath of the sky. Most of the data were taken
with the beam center at a zenith angle of  $40^\circ$. The
antenna beam has very low, symmetric sidelobes, and is shielded
from ambient temperature emission from the earth and the balloon
by  an inner ground screen attached to the dewar;
with this screen in place, the measured sidelobe
response is below $-55$~dB for all angles greater than $50^\circ$
from  the beam axis. A fixed outer ground screen provides
additional rejection of ground radiation.

\section{DATA}

The instrument was launched from Fort Sumner, NM,  on December
10, 1995.   The radio command uplink became intermittent a few minutes after
the package reached float altitude, severely restricting the
program of measurements.  The data divide into three  segments
chronologically, according to the zenith angles of observation:
the first $40^\circ$ tip, the $60^\circ$ tip, and the second
$40^\circ$ tip.   Details of these data segments are given in
Table 1.  In the latter two data segments, the instrument observed the moon,
providing a crosscheck of the pointing, beamsize, and
gain. Continuous in-flight gain calibration was accomplished by
alternating  the temperature of the reference load, $T_{ref}$, between
2.749~K and  4.091~K at twice the switch period. The temperature
difference between the colder of the two setpoints   and the sky
was $\lap 20$~mK, so that the effect of gain errors in
determining the sky temperature is negligible. Table 1 includes
raw numbers for the temperature of the sky signal,  with no
corrections for any systematic or instrument effects.   

\section{SYSTEMATIC EFFECTS}

Five sources of extraneous signal  are described here with
estimates of the errors in removing those sources.   The results
are summarized in Table 2.  The commanding difficulties precluded
planned in-flight measurements of several of these systematic
effects, so that calculations are used instead. However, the
experiment design keeps the size of all the corrections small
enough to allow for a precise determination of~\tcbr.

\subsection{Radiometer Offset}

The radiometer operation has been studied with multiple  separate
cooldown tests in which the sky horn is removed from the dewar
and replaced with a second thermally regulated  waveguide load,
identical in construction to the reference load. In particular,
two such calibration tests took place in Fort Sumner just before
the launch. Flight conditions are approximated during these tests
by pumping on the helium bath.

If the radiometer were ideal, when the measured temperatures of
the two waveguide loads were equal, no differential signal would
be observed from the radiometer.  This radiometer instead
exhibits offsets of $T_{off}= 7$~mK and 10~mK  in channels 2 and 3
respectively.   Four independent measurements of these offsets
were made in four separate cooldowns which agree to the relative
accuracy of the measurements, about 2 mK. These measured offsets
are assumed to apply during the flight. However, since their
origin is not yet certain, {\it errors equal to the magnitudes of
the offsets are assigned to their removal.} 
   
\subsection{Emission and Reflection from Waveguide/Optics}
 
The emission expected from the corrugated horn at 5~K is
calculated according to the methods described in Clarricoats,
Olver \& Chong (1975).  The horn was fabricated as a single piece
to eliminate the  excess emission associated with joints.   
The effect of the short waveguide
transition coupling the horn to the  waveguide switch is included
in the calculation, yielding  $T_{horn}=7^{+7}_{-3}$~mK. The
errors are conservative estimates. The measured reflection
coefficent of the horn is $<-28.5$~dB and that of the window is
$<-45$~dB. Since the in-flight circulator load  
temperature was within 100~mK of
the sky and reference load temperatures, differential reflection
effects are negligible. The expected window emission, $1\pm
1$~mK, is interpolated from measurements at other frequencies
(\cite{Cha71}).

The ambient-temperature ($211\pm 2$~K) inner ground screen (IGS), 
which is fixed to the  dewar, is near enough to the beam to
contribute  $T_{IGS} = 3\pm 3$~mK to the signal.  The magnitude
of this effect was determined after the flight  by lining the
inner surface of the inner ground screen with  microwave absorber
and measuring this lining's contribution to the sky signal. These
measurements were made by alternately observing the sky with and
without the absorber lining while operating the radiometer  at
ambient temperature to obviate the need for a thick emissive
window on the dewar. These results were scaled to account for the
much lower emissivity of the aluminum relative to the microwave
absorber, and the reduced effective emissivity of the absorber at
low incidence angles.

\subsection{Ground and Balloon Emission}

When the beam's zenith angle changed from $40^\circ$ to
$60^\circ$ during the flight, the observed sky temperature
increased by $\Delta T_{tip} \approx 10$~mK. (See Table 1.) 
Ground calibration tests indicate that  the radiometer offset is
unaffected by tipping, allowing attribution of $\Delta T_{tip}$
to an increase in the total radiation entering the antenna. The
associated atmospheric increase is expected to be $1.6\pm 1.6$~mK
(see Section 4.4), while the change in astrophysical foreground emissions is
$-0.4 \pm 1.4$~mK (see Section 4.5).   The remaining increase in sky
temperature, $\Delta T_{gnd} \approx 9$~mK, is used to constrain
the magnitude of the ground radiation, $T_{gnd}$, entering the
antenna during the $40^\circ$ tip. The largest contribution to
$T_{gnd}$ comes from ground radiation diffracting over the front
edge of the ground  screen. Measured beam patterns of the horn
antenna with the IGS in place indicate that the received ground pickup
from over the front edge should increase by a factor of $\gap 100$
upon tipping from $40^\circ$ to $60^\circ$.   The small
value of $\Delta T_{gnd}$ therefore implies that  $T_{gnd}$ is
less than 1~mK.  Since calculations of diffracted ground signals
are notoriously difficult, these estimates have been checked with
further measurements on the ground. Prior to flight, direct
measurements were made of the system beam map with
both ground shields in place.  The entire gondola was tipped
relative to a distance source, with the dewar fixed at several 
different angles relative to the gondola.  
These spot measurements agreed with predictions.  Nonetheless, a
conservative limit of $T_{gnd} < 5$~mK is adopted. In principle
radiation from the ground might reflect off the balloon and into
the horn antenna, but since the balloon fills only 0.02~sr and
has small reflectivity, this effect is negligible.

\subsection{Atmospheric Emission}

Extrapolation of the model described by Danese \& Partridge
(1989) to the flight altitude of 25~km,  using the
standard reference atmosphere for January midlatitudes
($30^\circ$N, \cite{Jur85}) yields $T_{atm}=3$~mK at $40^\circ$
zenith angle.   The average altitude dropped by 1.5~km  from the
first $40^\circ$ tip to the second, providing an opportunity to
test the model.  The model predicts a 2~mK increase due to this
drop; the measured increase is $2\pm 1$~mK.

\subsection{Foreground Emission} 

The principal celestial foreground emitters are the moon and
the  Galaxy; the data are also corrected  for the dipole of the
CMBR, as observed by COBE (\cite{Fix96}).  The dipole effect
varies from $-3$~mK to $+2$~mK across the region of sky observed.  
The Galactic signal is extrapolated from the continuum data at
408 MHz (\cite{Has70};  \cite{Has74}; \cite{Has81}) using a spectral
index of $2.8\pm 0.1$.  The Galactic signal  varies from 2~mK to
7~mK in these data. The Galactic signal is removed after binning
the data into sky coordinates. 

After the position-dependent sum of the dipole and Galactic
signals is subtracted from the data, the sky temperature data for
the second two data segments are
plotted versus the angle subtended from the beam center to the
moon's position. A Gaussian beam profile is then fit to these
data, reconfirming the measured beamwidth of $15^\circ$~FWHM and
implying an emission temperature for the moon of 240~K with the
moon about three-quarters full. This fit is used to correct the
data for the moon's signal.  All data for which the moon signal
is in excess of 10 mK (7\% of the data) are excluded from the determination
of~\tcbr.  

The average sum of these three foregrounds over the two
$40^\circ$ tips is $T_{fgnd} = 2\pm 2$~mK.

\section{RESULTS AND CONCLUSIONS}
\nopagebreak

The data and their systematic errors are summarized in Table 2.
The final result, averaged over the two bands, is a measurement
of~\tcbr$=2.730\pm 0.014$~K at 10.7 GHz, a measure accurate to
half a percent, made without  recourse to space missions.  It is
in agreement with a previous measurement of~\tcbr$=2.61\pm
0.06$~K at 10~GHz (\cite{Smo87}), and with the extremely precise  FIRAS
measurement of~\tcbr$=2.728\pm 0.004$~K (\cite{Fix96}).   
Several more measurements of \tcbr\ with accuracies
comparable to the results described here
will significantly tighten the constraints on $\mu$ and $Y_{ff}$.

\acknowledgments

We thank Skip Johnson, Ted Griffiths, Bill Dix, Laszlo Varga, Al
Dietrich  and Glenn Atkinson  for help in mechanical design and
construction.    We are grateful to  Richard Bradley and the NRAO
for supplying the HEMT amplifer. We express our gratitude to the NSBF
staff in Palestine and Fort Sumner for the excellent support 
before,  during and after the launch of the balloon, particularly
Mark Cobble.

This work was supported by NSF grant \# PHY89-21378.
Additional support was provided by NASA through Hubble Fellowship
grant \#HF-01063.01-94A awarded by STScI, which is operated by
the Association of Universities for Research in Astronomy, Inc.,
for NASA under contract NAS 5-26555.

\clearpage

\figcaption{Side view of the gondola with the beam's zenith angle
at 40\arcdeg, with an inset schematic of the radiometer. 
Windowscreen constitutes the outer ground screen, the sides of
which make an angle of 30\arcdeg\ to the vertical.  The inner	
ground screen is rolled aluminum.  The wooden deck is covered	
with aluminum.  The gondola is suspended from the balloon's	
flight train by the stainless steel cables depicted.  Two of the	
radiometer channels are selected via tunable filters. 	
Room-temperature isolators have been omitted for clarity.}

\clearpage 

\begin{deluxetable}{ccccc}
\tablecaption{DESCRIPTION OF THE THREE DATA SEGMENTS.\label{ta:data}}
\tablewidth{0pt}
\tablehead{
\colhead{Zenith Angle} & \colhead{GMT Range} & \colhead{Average
Altitude} & \colhead{Sky Signal} & \colhead{Sky Signal}\nl
\colhead{} & \colhead{(hrs)} &\colhead{(km)} & 
\colhead{Ch 2 (mK)} & \colhead{Ch 3 (mK)}
}
\startdata
First 40\arcdeg & 2.78--4.16  & 25.3 & 2738 & 2735\nl
60\arcdeg & 4.34--4.84 & 28.6 & 2749 & 2746\nl
Second 40\arcdeg & 5.51--6.54 & 23.8 & 2741 & 2739\nl
\enddata
\tablecomments{The gondola rotates 360\arcdeg\ in azimuth about once a minute.
The signal levels shown for Channels 2 and 3 are the sums of the raw measured
temperature differences between the sky and the reference load,
with the reference load temperature $2749\pm 4$~mK.  No
corrections for any systematic effects have been applied.
The statistical errors for each entry are $\lap 0.1$~mK.}
\end{deluxetable}

\begin{deluxetable}{lcc}
\tablecaption{MEASURED SIGNALS AND SYSTEMATIC ERRORS\label{ta:final}}
\tablewidth{360pt}
\tablehead{
\colhead{Parameter} & \colhead{Channel 2} & \colhead{Channel 3}\nl
\colhead{} & \colhead{(mK)} & \colhead{(mK)}
}
\startdata
\phs$T_{ref}$ & 	    \phs$2749\pm 4$ &		\phs$2749\pm 4$ \nl
\phs$\Delta T_{meas}$ & \phn\phn$-10 \pm 1$  & 	\phn\phn$-12\pm 1$\nl
$-T_{off}$ & 	    \phs\phn\phn\phn$7\pm 7$ & 	\phs\phn\phn$10\pm 10$\nl
$-T_{horn}$ & 	    \phn\phn\phn$-$7\,$^{\scriptsize +~3}_{\scriptsize -~7}$ 
	    &	    \phn\phn\phn$-$7\,$^{\scriptsize +~3}_{\scriptsize -~7}$\nl
$-T_{win}$ & 	    \phn\phn\phn$-1\pm 1$ & 	\phn\phn\phn$-1\pm 1$\nl
$-T_{IGS}$ & 	    \phn\phn\phn$-3\pm 3$ & 	\phn\phn\phn$-3\pm 3$\nl
$-T_{gnd}$ & 	    \phs\phn\phn\phn$>-5$ &     \phs\phn\phn\phn$>-5$\nl
$-T_{atm}$ & 	    \phn\phn\phn$-3\pm 3$ & 	\phn\phn\phn$-3\pm 3$\nl
$-T_{fgnd}$ 	 &  \phn\phn\phn$-2\pm 2$ & 	\phn\phn\phn$-2\pm 2$\nl
\phs\tcbr	& 	    \phs2730\,$^{\scriptsize +~10}_{\scriptsize -~13}$  
		&   \phs2731\,$^{\scriptsize +~12}_{\scriptsize -~15}$ \nl 
\enddata
\tablecomments{\tcbr\ in each channel is the sum of all the rows.  The final
error is obtained by summing the constituent errors in
quadrature.  $\Delta T_{meas}$ is the measured difference between the reference
load and the sky signal for all the data at zenith angle
40\arcdeg;  the error includes statistical and gain effects. $T_{IGS}$ denotes 
emission from the inner ground screen. 
The foreground term includes effects from both the Galaxy (with an average 
value of 3~mK) and the CMBR dipole (which decreases the signal by about 1~mK).  
Channel 2 spans the range 10.68--10.70~GHz, while
Channel 3 ranges from 10.6~GHz to 10.65~GHz. 
}
\end{deluxetable}

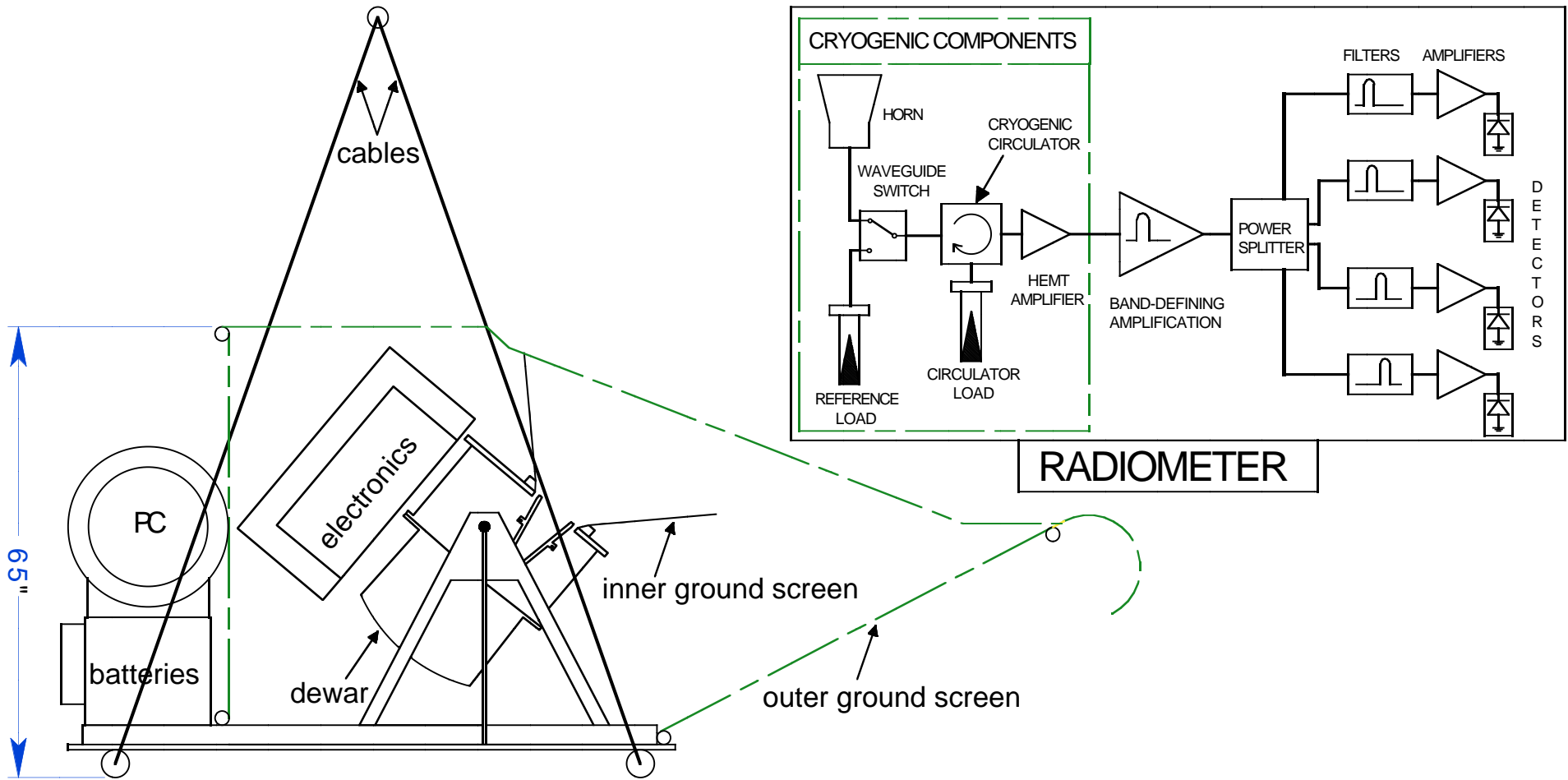
\end{document}